\documentclass[10pt,letterpaper]{article}
\usepackage[top=0.85in,left=2.75in,footskip=0.75in]{geometry}

\usepackage{amsmath,amssymb}

\usepackage{changepage}

\usepackage[utf8x]{inputenc}

\usepackage{textcomp,marvosym}

\usepackage{cite}

\usepackage{nameref,hyperref}

\usepackage[right]{lineno}

\usepackage{microtype}
\DisableLigatures[f]{encoding = *, family = * }

\usepackage[table]{xcolor}

\usepackage{array}

\newcolumntype{+}{!{\vrule width 2pt}}

\newlength\savedwidth

\newcommand\thickhline{\noalign{\global\savedwidth\arrayrulewidth\global\arrayrulewidth 2pt}%
\hline
\noalign{\global\arrayrulewidth\savedwidth}}

\usepackage{multirow}
\raggedright
\usepackage[aboveskip=1pt,labelfont=bf,labelsep=period,justification=raggedright,singlelinecheck=off]{caption}

\makeatletter
\renewcommand{\@biblabel}[1]{\quad#1.}
\makeatother

\usepackage{lastpage,fancyhdr,graphicx}
\usepackage{epstopdf}
\fancyheadoffset[L]{2.25in}
\fancyfootoffset[L]{2.25in}
\begin{document}
\vspace*{0.2in}

\begin{flushleft}
{\Large
\textbf\newline{Learning to Forecast and Forecasting to Learn from the COVID-19 Pandemic} % Please use "sentence case" for title and headings (capitalize only the first word in a title (or heading), the first word in a subtitle (or subheading), and any proper nouns).
}
\newline
\\
Ajitesh Srivastava\textsuperscript{1},
Viktor K. Prasanna\textsuperscript{1}
% Name3 Surname\textsuperscript{2,3\textcurrency},
% Name4 Surname\textsuperscript{2},
% Name5 Surname\textsuperscript{2\ddag},
% Name6 Surname\textsuperscript{2\ddag},
% Name7 Surname\textsuperscript{1,2,3*},
% with the Lorem Ipsum Consortium\textsuperscript{\textpilcrow}
\\
\bigskip
\textbf{1} 
Ming Hsieh Department of Electrical and Computer Enginnering, University of Southern California, Los Angeles, California, USA
\\
% \textbf{2} Affiliation Dept/Program/Center, Institution Name, City, State, Country
% \\
% \textbf{3} Affiliation Dept/Program/Center, Institution Name, City, State, Country
% \\
% \bigskip

% % Insert additional author notes using the symbols described below. Insert symbol callouts after author names as necessary.
% % 
% % Remove or comment out the author notes below if they aren't used.
% %
% % Primary Equal Contribution Note
% \Yinyang These authors contributed equally to this work.

% % Additional Equal Contribution Note
% % Also use this double-dagger symbol for special authorship notes, such as senior authorship.
% \ddag These authors also contributed equally to this work.

% % Current address notes
% \textcurrency Current Address: Dept/Program/Center, Institution Name, City, State, Country % change symbol to "\textcurrency a" if more than one current address note
% % \textcurrency b Insert second current address 
% % \textcurrency c Insert third current address

% % Deceased author note
% \dag Deceased

% % Group/Consortium Author Note
% \textpilcrow Membership list can be found in the Acknowledgments section.

% % Use the asterisk to denote corresponding authorship and provide email address in note below.
\bigskip
\{ajiteshs, prasanna\}@usc.edu

\end{flushleft}
\section*{Abstract}
Accurate forecasts of COVID-19 is central to resource management and building strategies to deal with the epidemic. We propose a heterogeneous infection rate model with human mobility for epidemic modeling, a preliminary version of which we have successfully used during DARPA Grand Challenge 2014. By linearizing the model and using weighted least squares, our model is able to quickly adapt to changing trends and provide extremely accurate predictions of confirmed cases at the level of countries and states of the United States. We show that during the earlier part of the epidemic, using travel data increases the predictions. Training the model to forecast also enables learning characteristics of the epidemic. In particular, we show that changes in model parameters over time can help us quantify how well a state or a country has responded to the epidemic. The variations in parameters also allow us to forecast different scenarios such as what would happen if we were to disregard social distancing suggestions.

% % Please keep the Author Summary between 150 and 200 words
% % Use first person. PLOS ONE authors please skip this step. 
% % Author Summary not valid for PLOS ONE submissions.   
% \section*{Author summary}
% Lorem ipsum dolor sit amet, consectetur adipiscing elit. Curabitur eget porta erat. Morbi consectetur est vel gravida pretium. Suspendisse ut dui eu ante cursus gravida non sed sem. Nullam sapien tellus, commodo id velit id, eleifend volutpat quam. Phasellus mauris velit, dapibus finibus elementum vel, pulvinar non tellus. Nunc pellentesque pretium diam, quis maximus dolor faucibus id. Nunc convallis sodales ante, ut ullamcorper est egestas vitae. Nam sit amet enim ultrices, ultrices elit pulvinar, volutpat risus.

%\linenumbers

% Use "Eq" instead of "Equation" for equation citations.
\section*{Introduction}
% Lorem ipsum dolor sit~\cite{bib1} amet, consectetur adipiscing elit. Curabitur eget porta erat. Morbi consectetur est vel gravida pretium. Suspendisse ut dui eu ante cursus gravida non sed sem. Nullam Eq~(\ref{eq:schemeP}) sapien tellus, commodo id velit id, eleifend volutpat quam. Phasellus mauris velit, dapibus finibus elementum vel, pulvinar non tellus. Nunc pellentesque pretium diam, quis maximus dolor faucibus id.~\cite{bib2} Nunc convallis sodales ante, ut ullamcorper est egestas vitae. Nam sit amet enim ultrices, ultrices elit pulvinar, volutpat risus.

% \begin{eqnarray}
% \label{eq:schemeP}
% 	\mathrm{P_Y} = \underbrace{H(Y_n) - H(Y_n|\mathbf{V}^{Y}_{n})}_{S_Y} + \underbrace{H(Y_n|\mathbf{V}^{Y}_{n})- H(Y_n|\mathbf{V}^{X,Y}_{n})}_{T_{X\rightarrow Y}},
% \end{eqnarray}

The recent outbreak of COVID-19 and the world-wide panic surrounding it calls for urgent
measures to contain the epidemic. Predicting the speed and severity of infectious diseases like
COVID-19 and allocating medical resources appropriately is central to dealing with epidemics.
The role of data science in this issue was brought to light when in December 2013, the first cases
of Chikungunya virus appeared in the Americas. In 2014 DARPA announced a Grand Challenge~\cite{darpachanllenge}
to predict the spread of Chikungunya virus in 55 countries of the western hemisphere. Through
monthly predictions and reevaluation over seven months, DARPA announced
the top winners of the challenge which included our team~\cite{darpawinners}. 
%We were also awarded a special mention for our interactive tool that allows the user to explore the predictions and retrain models with new data. 
While many winning methods relied on manual adjustments for generating predictions, ours was a completely automated approach, thus more generalizable. 
However, training such a model with rapidly changing epidemic trends is difficult. Often epidemic models are trained through numerical solutions to differential equations~\cite{cintron2020estimation} or through Bayesian inference~\cite{lekone2006statistical,dukic2012tracking}. Instead, we transform the model into a linear system and train it using weighted least squares. The more recent data is more heavily weighted to adapt to rapidly changing trends. Further, we explore various hyper-parameter selection strategies to identify the best model.

Learning the model also enables understanding of the dynamics of the epidemic and how it has changed over time. We utilize the changing parameters to study how various countries and US states have responded to the epidemic. We do so by proposing two measures: (i) Contact Reduction Score that measure how much a region has reduced transmission; (ii) and Epidemic Reduction Score that measures how much reduction in confirmed cases a region has achieved compared to a hypothetical scenario where the trends had remained the same as a reference day in the past. Further it enables analysis of scenarios into the future, for instance, what would be the trend of the epidemic if social distancing orders were released.

While we perform our analysis at state and country level.
In the future, we plan to generate similar predictions at county and city level. Applying such machine learning-based models to a finer level (from countries to states/cities) and larger scale (more `regions' of the world) brings unique challenges in terms of unreported/noisy data and large number of model parameters, which will be explored in a future work. In this paper, we present some of our initial results on country and state-level predictions. We forecast number of reported cases for US states and for all the countries. We understand that majority of the infections are conjectured to be unreported~\cite{lau2020internationally}. We still believe that number of reported cases is a good indicator of the stress on the healthcare system. Henceforth, ``number of infected cases'' refers to the number of reported cases. In future work, we plan to incorporate modeling of unreported cases informed by various ongoing antibodies studies.
We are also developing an interactive customizable tool that can be used to perform predictions using our model. A preliminary version of the tool has been made publicly available\footnote{\url{https://jaminche.github.io/COVID-19/}}.

\section*{Methodology}
%\subsection*{Etiam eget sapien nibh}

% % For figure citations, please use "Fig" instead of "Figure".
% Nulla mi mi, Fig.~\ref{fig1} venenatis sed ipsum varius, volutpat euismod diam. Proin rutrum vel massa non gravida. Quisque tempor sem et dignissim rutrum. Lorem ipsum dolor sit amet, consectetur adipiscing elit. Morbi at justo vitae nulla elementum commodo eu id massa. In vitae diam ac augue semper tincidunt eu ut eros. Fusce fringilla erat porttitor lectus cursus, \nameref{S1_Video} vel sagittis arcu lobortis. Aliquam in enim semper, aliquam massa id, cursus neque. Praesent faucibus semper libero.

% % Place figure captions after the first paragraph in which they are cited.
% \begin{figure}[!h]
% \caption{{\bf Bold the figure title.}
% Figure caption text here, please use this space for the figure panel descriptions instead of using subfigure commands. A: Lorem ipsum dolor sit amet. B: Consectetur adipiscing elit.}
% \label{fig1}
% \end{figure}
\subsection*{Modeling}
\label{sec:model}

\begin{figure}
  \begin{center}
    \includegraphics[width=0.41\textwidth]{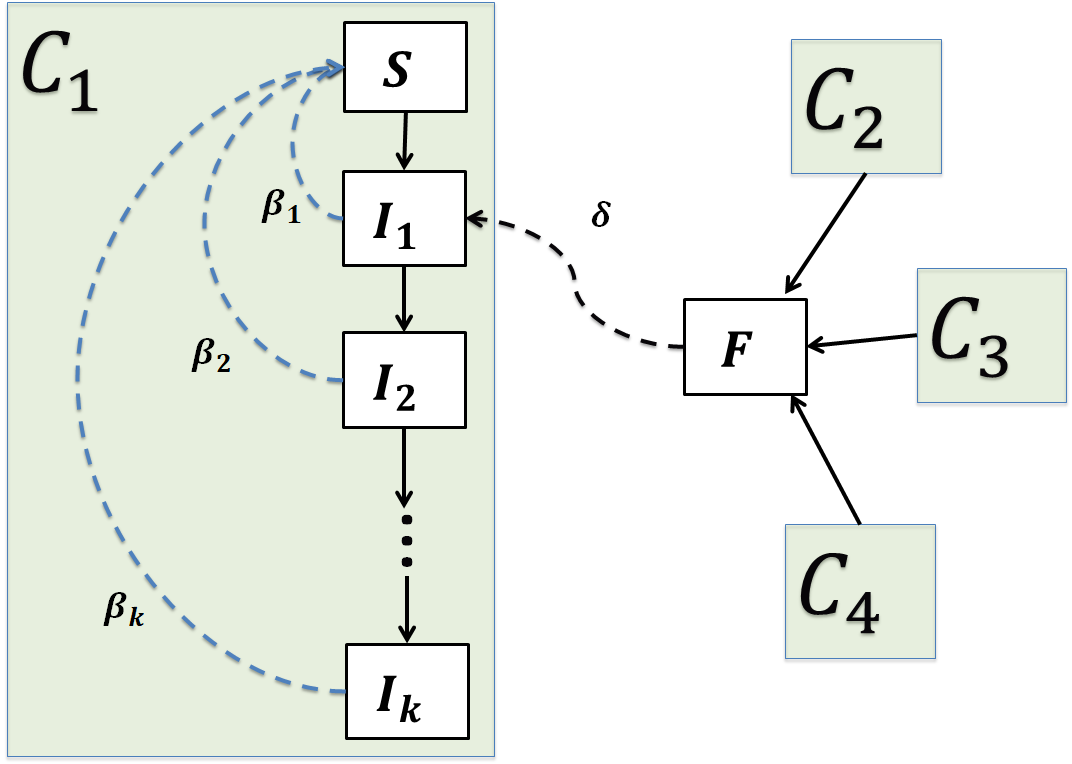}
  \end{center}
  \caption{\textbf{Heterogeneous Susceptible-Infected model with human mobility.}
  The model incorporates multiple infection sub-states with varying infection rates within a region and spread due to inter-region mobility.}
  \label{fig1}
\end{figure}

We propose an epidemic model (Fig.~\ref{fig1}) for spread of a virus like COVID-19 across the world which captures (i) temporally varying infection rates (ii) arbitrary regions, and (iii) human mobilty patterns. Within every region (hospital/city/state/country), an individual can exist in either one of two states: susceptible and infected. A susceptible individual gets infected when in contact with an infected individual at a rate depending on when that individual got infected, i.e., rate of infection  is $\beta_1$ for an individual infected at $t-1$, $\beta_2$ for an individual infected at $t-2$, and so on, thus resulting in $k$ sub-states of infection.
The hypothesis is that how actively one passes on the infection is affected by when they get infected. 
We assume that after being infected for a certain time, individuals no longer spread the infection, i.e., $\exists k$, such that $\beta_i = 0 \forall i>k$.
Also, people traveling from other regions can increase the number of infections in a given region. We assume that this infection can happen because of human mobility. Suppose $F(q, p)$ represents mobility from region $q$ to region $p$. 
%This includes estimates of (a) number of people flying from region $q$ to region $p$, and (b) number of people traveling between the regions by road.
Our model is represented by the following system of equations.

\begin{align}
\Delta S_t^p &= - \frac{S_{t-1}^p}{N^p} \sum_{i=1}^k \beta_i^p \Delta I_{t-i}^p \,,\label{eqn:delS}\\
\Delta I_t^p &= \frac{S_{t-1}^p}{N^p} \sum_{i=1}^k \beta_i^p \Delta I_{t-i}^p  + \delta \sum_q F(q, p) \frac{\sum_{i=1}^k \beta_i^q \Delta I_{t-i}^q}{N^q}\,. \label{eqn:delI}
\end{align}
Here, $S_t^p$ and $I_t^p$ represent the number of susceptible individuals and infected individuals respectively in the region $p$ at time $t$. 
%$\Delta S_t^p$ and $\Delta I_t^p$ are defined as $ S_t^p - S_{t-1}^p$ and $I_t^p - I_{t-1}^p$, respectively. 
Parameter $\delta$ captures the influence of passengers coming into the region.

To deal with the fact that using large values of $k$ may overfit the model and that data is likely to be noisy, we incorporate another hyperparameter $J$ in Eq~\ref{eqn:delI} as follows:
\begin{equation}
\Delta I_t^p = \frac{S_{t-1}^p}{N^p} \sum_{i=1}^k \beta_i^p (I_{t-iJ}^p -I_{t-(i-1)J}^p)  + \delta \sum_q F(q, p) \frac{\sum_{i=1}^k \beta_i^q (I_{t-iJ}^q -I_{t-(i-1)J}^q)}{N^q}\,. \label{eqn:delI_linear} 
\end{equation}
This creates a dependency on last $kJ$ days with $k$ parameters while having a ``smoothing'' effect due to combining infections over $J$ days. Note that if we set $k=1, J = \infty$, and ignore mobility ($\delta =  0$), this reduces to Suceptible-Infected (SI) model~\cite{zhou2006behaviors}. On the other hand, with bounded $k=1$ and $J < \infty$, the model is a variation of Suceptible-Infected-Released/Recovered (SIR) model~\cite{bjornstad2002dynamics}, where an infected individual is active for $J$ units of time.
While the model is applicable to any definition of regions, in this work, we focus on country-level and US state-level forecasts, only. Further, there is a lag between an individual contracting the virus and their case being reported. We do not account for this lag, assuming that it is constant.
County-level and city-level forecasting accounting for the lag and unreported cases is planned for future work. 

\subsection*{Training}

To train the model, we linearize it, by setting $\delta^p_i = \delta \beta_i^p$ and learning it as an independent parameter. This makes the model more general by allowing different infection rates for the travelers. This is different from traditional approach of fitting one curve to the data with fixed initial values, which is computationally expensive and cannot capture rapidly changing trends.
Let 
\begin{align}
\mathbf{\beta}^p &=  
\begin{bmatrix}
\beta_1^p & \dots & \beta_k^p & \delta_1^p & \dots & \delta_k^p
\end{bmatrix}
\\
\mbox{And, }    \mathbf{X}_t^p &= \begin{bmatrix}
           S_t(I_{t}^p -I_{t-J}^p) \\
           \vdots \\
           S_{t-(k-1)J}(I_{t-(k-1)J}^p -I_{t-kJ}^p) \\
           \sum_q \frac{F(q, p)}{N^q}(I_{t}^q -I_{t-J}^q)\\
           \vdots \\
           \sum_q \frac{F(q, p)}{N^q}(I_{t-(k-1)J}^q -I_{t-kJ}^q)]^T
         \end{bmatrix}
  \end{align}
Then, our model can be represented by the linear equation
\begin{equation}
    \Delta I_t^p = \mathbf{\beta}^p \mathbf{X}_t^p\,,
\end{equation}
which allows us to learn the parameters using a constrained linear solver. This formulation works only if $k$ and $J$ are same for all regions. To allow different hyper-parameters for different regions, we used a further simplification:
\begin{align}
\mathbf{\beta}^p &=  
\begin{bmatrix}
\beta_1^p & \dots & \beta_k^p & \delta^p
\end{bmatrix}
\\
\mbox{And, }    \mathbf{X}_t^p &= \begin{bmatrix}
           S_t(I_{t}^p -I_{t-J}^p) \\
           \vdots \\
           S_{t-(k-1)J}(I_{t-(k-1)J}^p -I_{t-kJ}^p) \\
           \sum_q \frac{F(q, p)}{N^q}(I_{t}^q -I_{t-kJ}^q)]^T\,.
         \end{bmatrix}
  \end{align}
To incorporate the fast evolving trend of COVID-19 due to changing policies, we use weighted least squared to learn parameters $\beta_i^p$ and $\delta_i^p$ from available reported data. The best fit in the least-squares sense minimizes the sum of squared weighted residuals, i.e., the difference between observed data and predicted values provided by our learned model. We incorporate a forgetting factor $\alpha \leq 1$ in our minimization to put more weight to the recent infection trend when learning the model. Lower $\alpha$ implies more emphasis on the more recent data. We compute the minimum of the sum of squares of weighted errors as
\begin{align}
    LSE &= \sum_{t=1}^{T} \alpha^{T-t}(\Delta \hat{I}_t^p - \Delta I_t^p)^2 \\
    &= \sum_{t=1}^{T} (\alpha^{\frac{T-t}{2}}\Delta\hat{I}_t^p - \alpha^{\frac{T-t}{2}}\ \mathbf{\beta}_p \mathbf{X}_t^p)^2\,,
\end{align}
where $\Delta \hat{I}_t^p$ is the actual number of newly infected individual.

\subsection*{Data}

We present our analysis on two datasets - (i) Global: country-level data with each country defined as a region; and (ii) US states: state-level data for Unites States with each state defined as a region. Country-level infections were derived from confirmed cases obtained from JHU CSSE COVID19 dataset~\cite{JHUdata}. Population of the countries were obtained from the World Bank dataset~\cite{country-travel}. Inter-country travel data were obtained from KCMD Global Transnational Mobility dataset~\cite{country-travel}, which makes the estimation on the basis of global statistics on tourism and air passenger traffic. Number of infections for US states were obtained from confirmed cases compiled by New York Times~\cite{USdata}. Population of each state was obtained from US Census Bureau~\cite{USpopu}. Inter-state travel was estimated from number of flights flying between airports in the United States~\cite{UStravel}.

% Results and Discussion can be combined.
\section*{Results}
The hyper-parameters to be picked include $k$, $J$, and $\alpha$. The best hyper-parameter set is identified by a grid search on $k, J,$ and $\alpha$, which minimizes the Root Mean Squared Error over a validation set of $H=3$:

\begin{equation}
    RMSE =  \sqrt{\frac{\sum_{t=T-H+1}^T (\Delta \hat{I}_t^p - \Delta I_t^p)^2}{H}}
\end{equation}
We pick $\alpha$ from $\{0.1, 0.2, \dots, 1.0\}$, while $k, J \in \{1, 2, \dots, 14\}$. We enforce $Jk \leq 14$ days so that the dependence of new infections is bounded by the previous two weeks. This is along the lines of the motivation for 14 days of quarantine\footnote{\url{https://www.cdc.gov/coronavirus/2019-ncov/travelers/after-travel-precautions.html}}.
We experiment with two methods of setting the hyper-parameters: (i) fixed - same set of hyper-parameters for all countries/states (ii) variable - specialized parameters for each state/country. The results are presented next.
\subsection*{Test Results}

We evaluated the results for a horizon $H=3$ with two hyper-parameters selection schemes termed SI-kJ$\alpha$ (variable) and SI-kJ$\alpha$ (fixed). The best heper-parameters were selected by further splitting the training data, to set the last three days for validation. We compare these results against a recent generalized version of SEIR model~\cite{peng2020epidemic}. Besides RMSE, we also measure the mean absolute percentage error given by
\begin{equation}
    MAPE = \frac{1}{H}\sum_{t=T-H+1}^T \frac{|\Delta \hat{I}_t^p - \Delta I_t^p|}{\Delta \hat{I}_t^p}
\end{equation}
The comparison is shown in Tab.~\ref{table1}. We used a publicly available implementation of the baseline~\footnote{\url{https://github.com/ECheynet/SEIR}}. We observe that both fixed and variable hyper-parameter selections widely outperformed the baseline. Performances of both fixed and variable scheme were comparable. We use the average of both predictions as an `ensemble' prediction which performed better than both schemes, individually. It should be noted that the number of cases across different countries has orders of magnitude difference. Filtering out countries with small number of cases will significantly reduce MAPE but increases RMSE. As an example, considering only US for country-level prediction, the RMSE for all three (fixed, variable, and ensemble) is 6886.9, which translates to 1.12\% MAPE. All the code used in our experiments are available on Github \footnote{\url{https://github.com/scc-usc/ReCOVER-COVID-19}}.

\begin{table}[!ht]
\begin{adjustwidth}{-2.25in}{0in} % Comment out/remove adjustwidth environment if table fits in text column.
\centering
\caption{
{\bf Comparison of performance}}
%\begin{tabular}{|l+l|l|l|l|l|l|l|}
\begin{tabular}{|l|cc|cc|}
\hline
%\multicolumn{4}{|l|}{\bf Heading1} & \multicolumn{4}{|l|}{\bf Heading2}\\ \thickhline
Method & RMSE (US) & MAPE (US) & RMSE (Global) & MAPE (Global)\\ \thickhline
SI-kJ$\alpha$ (variable) & 333.3 & 6.82\% & 462.6 & 13.64\% \\
SI-kJ$\alpha$ (fixed) & 342.05 & 6.58\% & 456.0 & \textbf{11.22}\% \\
SI-kJ$\alpha$ (ensemble) & \textbf{316.3} & \textbf{5.93}\%  & \textbf{355.9} & 11.37\% \\
Gen-SEIR \cite{peng2020epidemic} & 2106.4 & 14.31\%  &  7471.2* & 41.06\%* \\ \hline
\end{tabular}
\begin{flushleft} Our approach with variable hyper-parameter set has the best performance in terms of both MAPE and RMSE. \\ *For some countries, the baseline method could not converge and for some the produced MAPE was greater than 100\%. We ignore those countries to favor the baseline which results in averaging over 147 countries.
\end{flushleft}
\label{table1}
\end{adjustwidth}
\end{table}

\begin{figure}[!ht]
\begin{adjustwidth}{-2.25in}{0in} % Comment out/remove adjustwidth environment if table fits in text column.
    \centering
    \includegraphics[width = 1.2\textwidth]{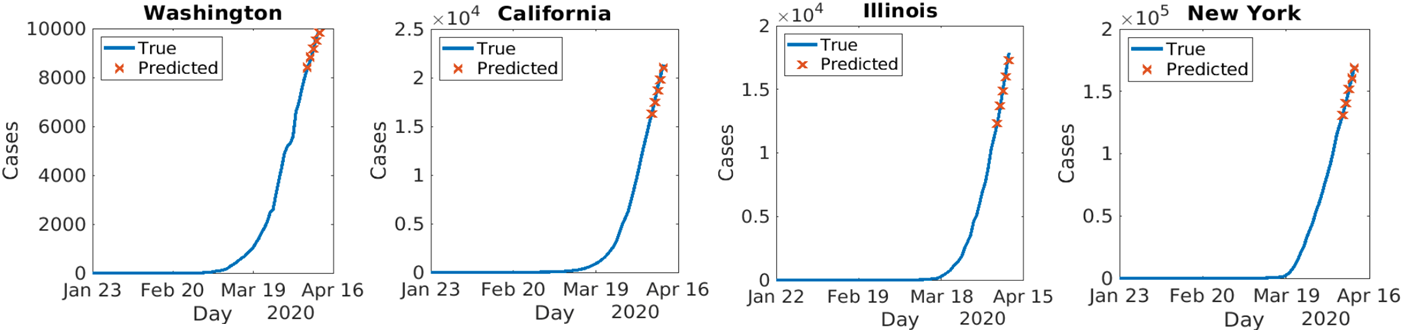}
    \caption{\textbf{Evaluation on US states.} The model accurately predicts on the last 5 days not included in the training. Four states are selected arbitrarily for illustration.}
    \label{fig2}
    \end{adjustwidth}
\end{figure}

\begin{figure}[!ht]
\begin{adjustwidth}{-2.25in}{0in} % Comment out/remove adjustwidth environment if table fits in text column.
    \centering
    \includegraphics[width = 1.2\textwidth]{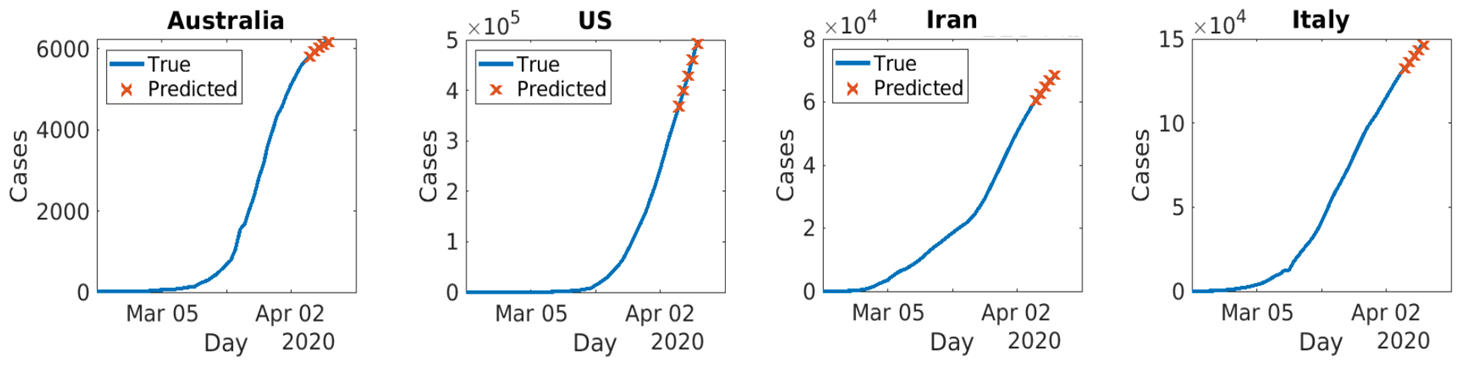}
    \caption{\textbf{Evaluation on countries.} The model accurately predicts on the last 5 days not included in the training. Four countries are selected arbitrarily for illustration.}
    \label{fig3}
    \end{adjustwidth}
\end{figure}

Fig.~\ref{fig2} shows the test results using SI-kJ$\alpha$ (ensemble) prediction on $H=5$ days for four US states. These were arbitrarily selected; other states show similar results as well. 
Fig.~\ref{fig3} shows the test results on $H=5$ days for four countries. Again, these were arbitrarily selected. Note that the predictions are extremely accurate. For these results, we set all mobility terms ($F(p, q)$) to zero to reflect that due to ``stay-at-home" policy around the world, the recent spread due to travel is negligible. This also reduces the number of parameters to be learned. For an earlier date, we include mobility and discuss the details next.

%We also tested the fit with the mobility term included, and trained the original non-linear formulation using gradient descent. It resulted in an extremely small , suggesting that our model was able to identify that mobility played negligible role in the recent data. Regardless, we set it to zero, and train the linearized version to avoid overfitting with too many parameters. 


% \subsubsection*{Effect of Mobility}
% While we set the mobility parameters to zero for recent predictions, we have scenarios from the prior days data where adding mobility parameters improve the predictions. Particularly, for confirmed cases in Louisiana, the best fit with travel produced highly accurate forecasts for end of March. On the other hand, predictions that does not include travel from out of state result in poor fit as shown in Fig.~\ref{}.

% \begin{figure}
%     \centering
%     \includegraphics[width = 0.6\textwidth]{./Louisiana}
%     \caption{Caption}
%     \label{fig:my_label}
% \end{figure}

\subsection*{Effect of Travel}
To measure the effect of travel, we train our model with data until March 18th, with and without travel data. The next three days are used as validation set to identify best hyper-parameters and the following three days are used as test set for evaluation. Table~\ref{table2} shows validation and test errors for both variable and fixed hyperparameter selection scheme, each with and without travel. Observe that in all cases including travel data reduces RMSE in the test set. For US states, variable hyper-parameter scheme performs better, while for Global, fixed scheme has a lower RMSE. This may be due to lack of enough data resulting in overfitting of variable hyperparameter scheme. This is evident from the fact that the validation errors are much lower than the test errors. This observation also supports the decision of going forward with the ensemble approach instead of choosing one of variable or fixed schemes.
\begin{table}[!ht]
\begin{adjustwidth}{-2.25in}{0in} % Comment out/remove adjustwidth environment if table fits in text column.
\centering
\caption{
{\bf Inclusion of Travel Data}}
%\begin{tabular}{|l+l|l|l|l|l|l|l|}
\begin{tabular}{|l|l|cc|cc|}
\hline
%\multicolumn{4}{|l|}{\bf Heading1} & \multicolumn{4}{|l|}{\bf Heading2}\\ \thickhline
& &  \multicolumn{2}{c|}{US} & \multicolumn{2}{c|}{Global}\\
& Method & RMSE & MAPE & RMSE & MAPE\\ \thickhline
\multirow{4}{*}{Validation} & travel, variable & \textbf{25.6} & 10.11\%  &  78.6 & 9.39\% \\
& without travel, variable & 26.9 & \textbf{9.71}\% & \textbf{74.4} & \textbf{8.65}\% \\
& travel, fixed & 48.3 & 21.94\%  & 138.5 & 22.61\% \\
& without travel, fixed & 49.1 & 23.61\% & 139.38 & 25.51\% \\
\hline
\multirow{4}{*}{Test} & travel, variable  & \textbf{147.3} & 19.93\%  &  248.4 & 21.353\% \\
& without travel, variable & 166.7 & \textbf{18.51}\% & 348.2 & 23.15\% \\
& travel, fixed & 207.0 & 25.08\%  & \textbf{242.6} & \textbf{19.50}\% \\
& without travel, fixed & 186.6 & 19.52\% & 286.8 & 21.42\% \\
\hline
% \multirow{4}{*}{Test (1-100)} & travel, variable  & \textbf{130.5} & 19.73\%  &  41.1 & 28.11\% \\
% & without travel, variable & 166.7 & \textbf{18.51}\% & 45.5 & 30.71\% \\
% & travel, fixed & 192.1 & 24.89\%  & \textbf{36.6} & \textbf{23.13}\% \\
% & without travel, fixed & 203.4 & 20.47\% & 44.3 & 29.89\% \\
% \hline
\end{tabular}
\begin{flushleft} Test errors suggest that including travel reduces RMSE.
\end{flushleft}
\label{table2}
\end{adjustwidth}
\end{table}

\subsection*{Insights from Parameters}
%The best performing fixed hyper-parameters were $k=2, J=5, \alpha = 0.2$, suggesting that the dependence of new infections is best explained by the prior $k\times J = 10$ days. The low value of $\alpha$ reflects the need to quickly update to more recent data due to rapidly changing infection trends. This observation closely matches with the best hyper-parameters obtained from US state-wise training, where the median value of $k\times J$ was 8. Note that these values may change with time as the epidemic is affected by our response.

Since the speed of infection is driven by $\beta_i^p$, we can assess the effect of a region's effort to battle COVID-19 by the change observed in these parameters. 
We measure the number of transmissions a susceptible person receives from an infected individual, assuming the infections are uniformly distributed across all sub-states. Another approach to define this quantity is to measure the number of new infections, given that past infections are uniformly increasing. This allows us to quantify a measure of ``contact'' without relying on the state of individuals (infected/susceptible) in the population as 
\begin{equation}
    \tau_N^p \propto \sum_i J^p \beta_i^p\,.
\end{equation}
We define Contact Reduction Score (CRS) for a region (country/state) as the fractional change in this number of transmissions:
\begin{equation}
    CRS(p) = \frac{ \tau_N^p(old) - \tau_N^p(new) }{\tau_N^p(old)}\,.
\end{equation}
We picked a date in the middle of March as the reference day to learn $\tau_N^p(old)$, and compared against $ \tau_N^p(new)$ obtained from training up to April 10. Since, the ensemble approach for hyper-parameters selection has the best performance, we compute the number of transmissions as the average of transmissions obtained from both scenarios. 
Further, we also define Epidemic Reduction Score, which measures the fractional reduction in number of infections compared to the scenario where the trend of infection at reference day had continued. 

\begin{figure}
\begin{adjustwidth}{-2.25in}{0in} % Comment out/remove adjustwidth environment if table fits in text column.
    \centering
    \includegraphics[width=1.3\textwidth]{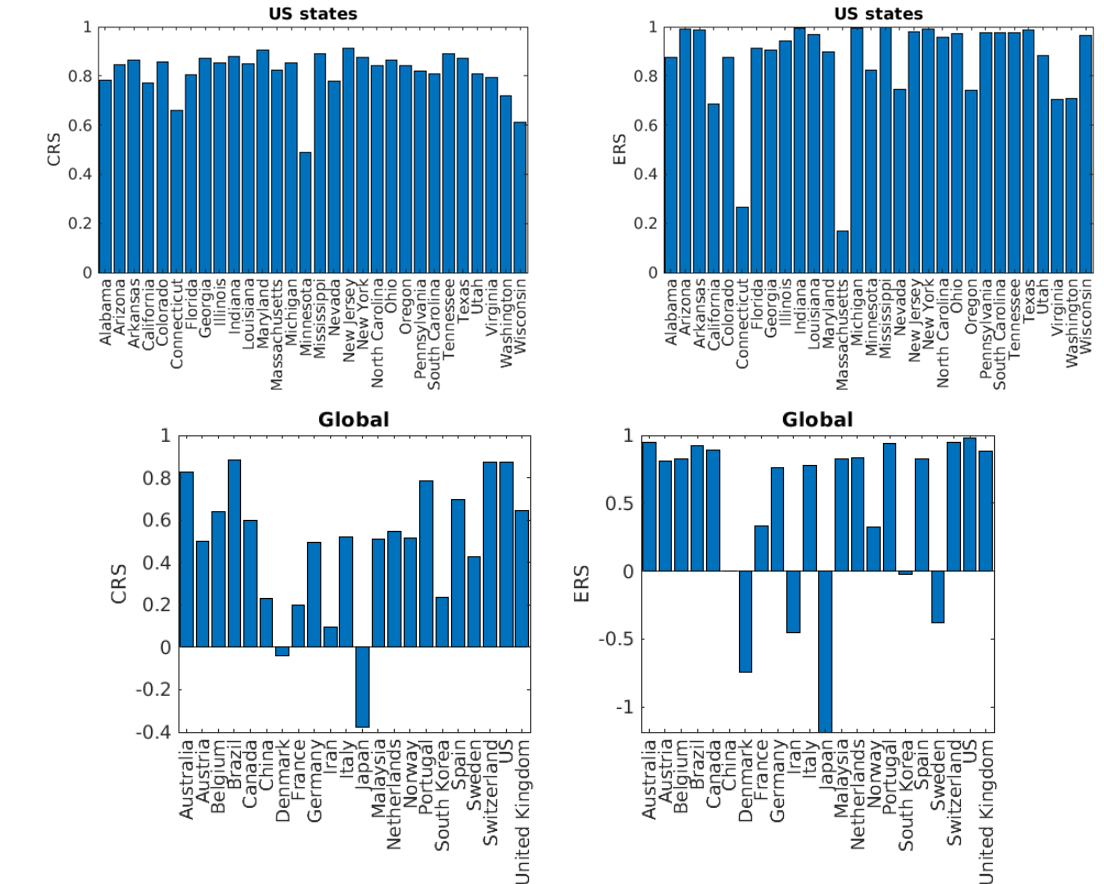}
    \caption{\textbf{Epidemic response evaluation.} Contact Reduction Score (CRS) and Epidemic Reduction Score (ERS) for various US states and countries evaluated for April 10th with March 21st as the reference day.}
    \label{fig4}
    \end{adjustwidth}
\end{figure}

The Reduction Scores for US states and global countries are shown in Fig.~\ref{fig4}. We ignored all the regions that had less than $100$ cases for US state-level and $1000$ cases from country-level data on the reference day in order to make the comparison reliable. 

\begin{figure}
\begin{adjustwidth}{-2.25in}{0in} % Comment out/remove adjustwidth environment if table fits in text column.
    \centering
    \includegraphics[width=\textwidth]{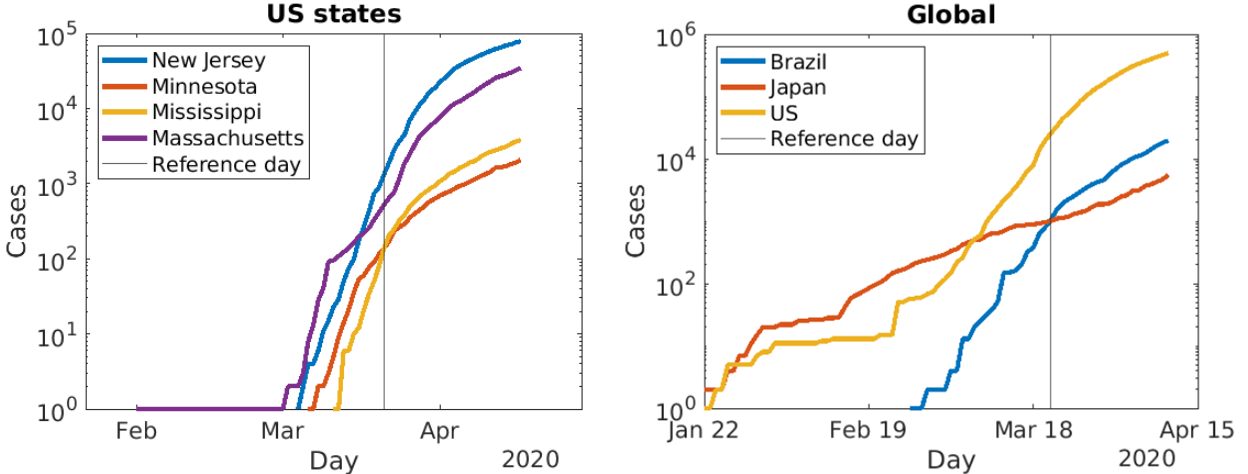}
    \caption{\textbf{Infection trends.} Comparison of states and countries confirmed cases over time. Vertical line represents the reference day with respect to which the reduction scores have been computed. Please note that the y-axis is logarithmic.}
    \label{fig5}
    \end{adjustwidth}
\end{figure}

Among the $30$ US states that qualify based on the threshold of $100$ cases at baseline, the state with best CRS was New Jersey, and Minnesota scored the worst. Mississippi has the top ERS with Massachusetts at the bottom of the list. Mississippi is also close to New Jersey in CRS. 
We note that Mississippi has been awarded between C and F since the reference day to early April for its ``Social Distancing'' score by Unacast ~\footnote{https://www.unacast.com/covid19/social-distancing-scoreboard}. This suggests that while ``Social Distancing'' score captures percentage reduction in average mobility, it does not provide the complete picture of the changes in infection dynamics. A visual inspection reveals that Mississippi has indeed shown significant change (Fig.~\ref{fig5}) compared to reference day when number of infections were increasing much more rapidly. At country-level, Brazil has the best CRS with Japan at the bottom, among $22$ countries selected based on a threshold of $1000$ infections at reference day. Based on ERS, US tops the list with Japan at the bottom again. We emphasize that these rankings are sensitive to the reference day.

While both CRS and ERS measure a region's response to the epidemic, we believe that CRS is a better metric to evaluate a region's efforts. This is due to the fact that CRS only depends on the model parameters that can be controlled by limiting contact with others and other policies to reduce transmission. On the other hand, ERS depends on number of current/past infections which are not completely controlled by changing policies. For instance, a country which already has large number of cases (yet significantly less than its susceptible population) will experience a higher increase compared to a country with fewer cases, even when they have identical model parameters. 

\subsection*{Forecasts}
\begin{figure}[!ht]
\begin{adjustwidth}{-2.25in}{0in} % Comment out/remove adjustwidth environment if table fits in text column.
    \centering
    \includegraphics[width = 1.2\textwidth]{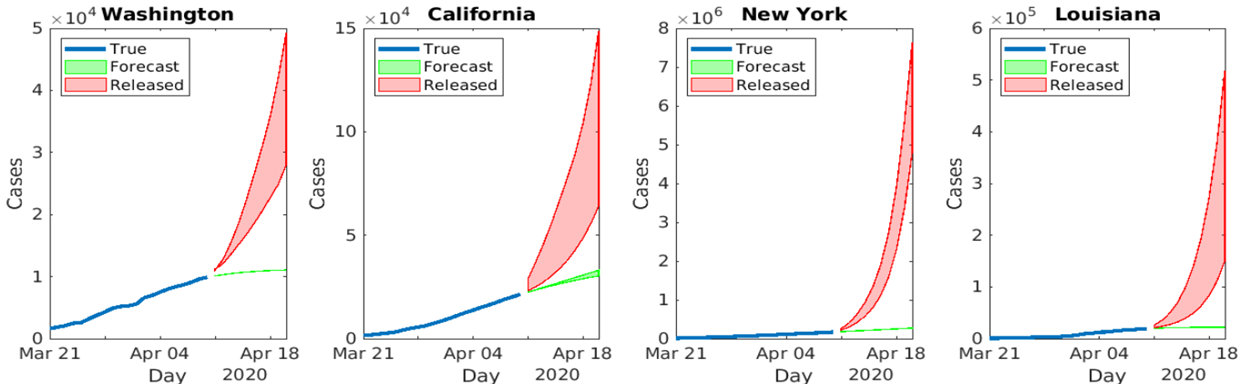}
    \caption{\textbf{Forecasts for US states.} The green region depicts the forecast according to the current trend. Red region represents the forecast for the scenario where the precautionary measures are instantly removed . Four states are selected arbitrarily for illustration.}
    \label{fig6}
    \end{adjustwidth}
\end{figure}

\begin{figure}[!ht]
\begin{adjustwidth}{-2.25in}{0in} % Comment out/remove adjustwidth environment if table fits in text column.
    \centering
    \includegraphics[width = 1.2\textwidth]{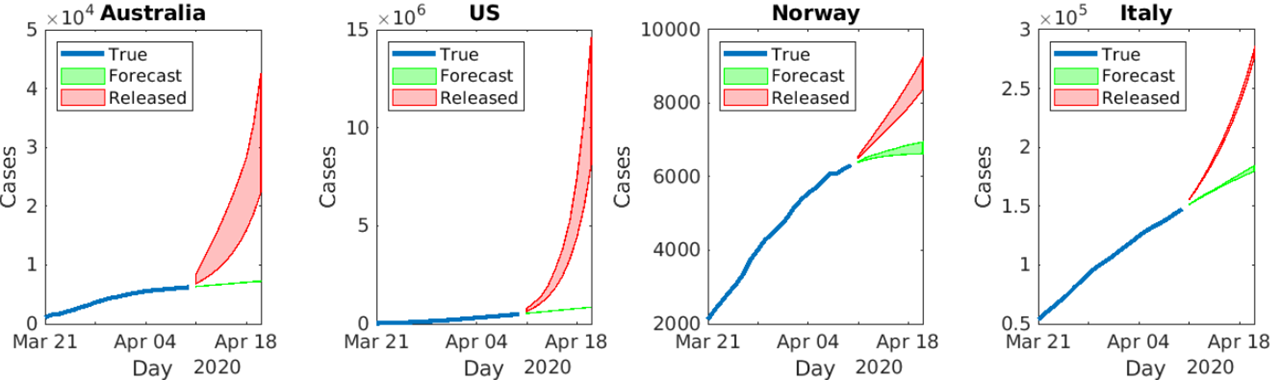}
    \caption{\textbf{Forecasts for countries.} The green region depicts the forecast according to the current trend. Red region represents the forecast for the scenario where the precautionary measures are instantly removed . Four countries are selected arbitrarily for illustration.}
    \label{fig7}
    \end{adjustwidth}
\end{figure}

We can use our models not only to generate forecasts, but also for emulating scenarios. As an example, we compute forecasts with two models: (i) a model trained on most recent data with no inter-region mobility, and (ii) a model trained on data until mid-March (reference day) including inter-region mobility. The reference day was set to middle of March as the majority of US states and countries had not actively imposed ``social distancing'' suggestion. The results with both models are shown in Figs.~\ref{fig6} and~\ref{fig7}. Instead of showing a single curve for forecasts, we show a range owing to the difference in fixed and variable hyper-parameter schemes. All plots suggest that, even though the current trend is leading to limited number of infections, immediately releasing all precautions can result in a rapid rise of the epidemic.

% \begin{figure}
%     \centering
%     \includegraphics[width = 0.9\textwidth]{./Georgia}
%     \caption{Caption}
%     \label{fig:my_label}
% \end{figure}

\section*{Discussion}
\subsection*{Hyper-parameter Selection} 
We have proposed two schemes for selecting hyper-parameters, fixed and variable. While both schemes have similar performance, variable scheme has a significant difference in test and validation error, suggesting that it is prone to over-fitting. On the other hand, with more data this issue may be resolved. While, currently our `ensemble' approach is simply the average of the predictions using both schemes, 
the ideal approach may be a hybrid one, where a collection of regions share the same hyper-parameters. This is especially useful when there are very few cases in a region. Then the hyper-parameters and even the parameters for this region can be set to be equal to that of a ``similar'' region from where more data is available. One way to identify such similar regions is by clustering regions based on their trends in a previous epidemic. In this way, active reporting for COVID-19 can be used to improve forecast of the next epidemic at an early stage. We will explore this in a future work.

\subsection*{Unreported Cases}
\begin{figure}[!ht]
\begin{adjustwidth}{-2.25in}{0in} % Comment out/remove adjustwidth environment if table fits in text column.
    \centering
    \includegraphics[width = 0.8\textwidth]{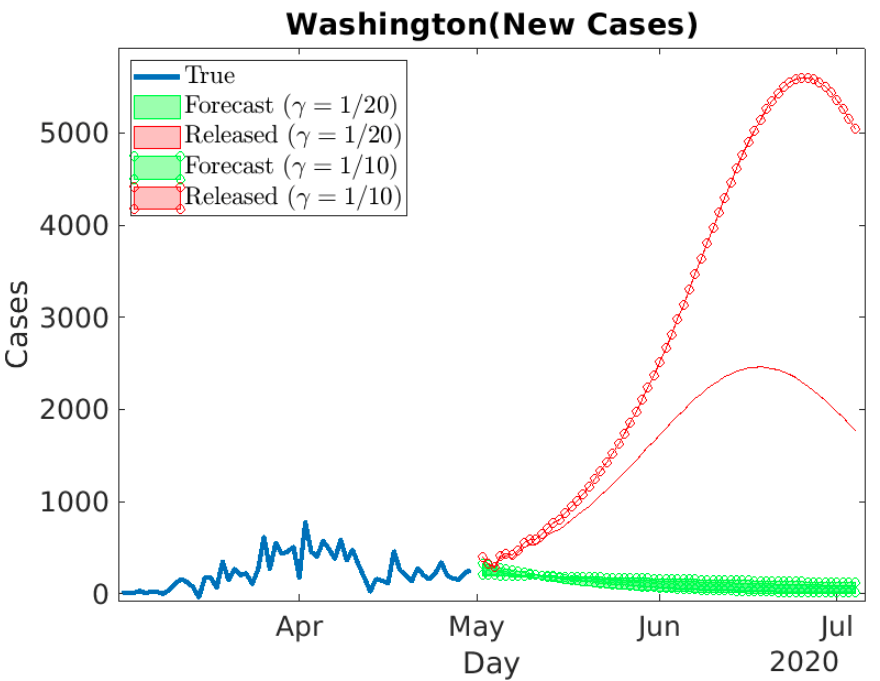}
    \caption{\textbf{Forecasts of reported cases considering total cases.} As in the previous figure, we show scenarios for forecast with the current trend and that with no precautionary measures. For each of those scenarios, the figure shows two possibilities of the fraction of reported cases with $\gamma = 1/20$ and $1/40$.}
    \label{fig8}
    \end{adjustwidth}
\end{figure}

We have only modeled reported cases, as they are an indicator of the stress on the healthcare system. However, unreported cases may affect the long term dynamics. The unreported cases can be classified into two categories: (i) unreported cases - those who get infected over the course of the epidemic but do not report it; and (ii) immune cases - those who have the antibodies without being infected during the epidemic. For unreported cases, we can add another state to our model: for an individual in the $i^{th}$ ``infected'' sub-state, a report will be made with probability $\gamma_i^p$. Thus, the total number of new reported cases is given by $\Delta R  = \sum_{i=1}^k \gamma_i^p \Delta I_{t-i}^p$. Then the parameters will be learned by fitting the reported cases to $R_t^p$. The immune cases can be modeled as considering them not-susceptible. Suppose, $\rho^p$ is the probability of a randomly selected individual in region $p$ to be immune. Then the number of susceptible individuals at time $t$ is given by $S_t^p = (1-\rho^p)N^p - I_t^p$. We can integrate these cases complemented by the ongoing various anti-body studies and surveys that will assist with more accurate learning of the parameters. Learning these parameters from the data without considering these studies is difficult. This is due to the fact that the number of total cases, even if it is 100 times the reported cases, is currently a very small fraction of the population. As of now, we allow $\gamma_i^p$ and $\rho^p$ to be inputs in our model. These can be used to study long-term scenarios with various inputs. For example, Fig.~\ref{fig8} shows the number of new positive cases (reported) for various values of $\gamma_i^p = \gamma$.

\section*{Conclusion}
We have proposed a heterogeneous infection rate model with human mobility to model the spread of COVID-19 at country and state-level. Our model incorporates a forgetting factor to quickly adapt to the rapidly changing trends due to the changing policies of how we respond to the epidemic. With a weighted least square training and identifying the right hyper-parameters, we are able to achieve highly accurate predictions. The parameters obtained over different time-intervals allow us to measure how various regions (states/countries) have responded to the epidemic. In particular, we have defined Contact Reduction Score and Epidemic Reduction Score that respectively measure reduction in transmission, and reduction in epidemic spread compared to a hypothetical scenario where a trend of a prior reference point were to continue. These varying parameters also represent different policies of the past, and hence allow us to simulate scenarios into the future. For instance, using the parameters learned up to the point before ``social distancing'' orders, we can forecast the trajectory of the epidemic, if everyone were to stop taking distancing precautions.

\section*{Acknowledgments}
This work was supported by National Science Foundation Award No. 2027007.
The authors would like to thank Frost Tianjian Xu for preparing datasets and Jamin Chen for integrating our methods into a web-based visualization. The authors also thank Prathik Rao and Kangmin Tan for implementing and testing various ML approaches.

\nolinenumbers

% Either type in your references using
% \begin{thebibliography}{}
% Text
% \end{thebibliography}
%
% or
%
% Compile your BiBTeX database using our plos2015.bst
% style file and paste the contents of your .bbl file
% here. See http://journals.plos.org/plosone/s/latex for 

\bibliography{refs}

% \begin{thebibliography}{10}

% \bibitem{bib1}
% Conant GC, Wolfe KH.
% \newblock {{T}urning a hobby into a job: how duplicated genes find new
%   functions}.
% \newblock Nat Rev Genet. 2008 Dec;9(12):938--950.

% \bibitem{bib2}
% Ohno S.
% \newblock Evolution by gene duplication.
% \newblock London: George Alien \& Unwin Ltd. Berlin, Heidelberg and New York:
%   Springer-Verlag.; 1970.

% \bibitem{bib3}
% Magwire MM, Bayer F, Webster CL, Cao C, Jiggins FM.
% \newblock {{S}uccessive increases in the resistance of {D}rosophila to viral
%   infection through a transposon insertion followed by a {D}uplication}.
% \newblock PLoS Genet. 2011 Oct;7(10):e1002337.

% \end{thebibliography}

\end{document}